\documentclass[aps, prx, a4paper, showpacs, twocolumn, english, 10pt, longbibliography]{revtex4-2}
\usepackage{bbm, amsthm, bm,textcomp, nicefrac,geometry,ragged2e}
\usepackage[dvipsnames]{xcolor}
\usepackage{float}
\usepackage[bbgreekl]{mathbbol}
\usepackage{graphicx,epstopdf,color,verbatim,enumitem,ulem}
\usepackage{pbox,array}
\usepackage{mathrsfs}
\usepackage{amssymb}
\usepackage{textgreek}
\usepackage{mathtools}
\usepackage{stackrel}
\usepackage[thinlines]{easytable}
\usepackage{amsmath}
\usepackage[utf8]{inputenc}
\makeatletter
\usepackage{soul}
\usepackage[caption=false]{subfig}
\usepackage{hyperref}
\hypersetup{
    colorlinks = true,
    linkcolor = Red,
    citecolor = Red,
    filecolor = Black,
    urlcolor = Black,
}
\usepackage[T1]{fontenc}
\usepackage[caption=false]{subfig}
\usepackage{babel}
\usepackage[linesnumbered,ruled]{algorithm2e}

\addto\captionsenglish{}

\definecolor{brickred}{rgb}{0.8, 0.0, 0.0}

\begin{document}

\title{Long-ranged gates in quantum computation architectures with limited connectivity}

\author{Wolfgang D\"ur}
\email[Corresponding author: ]{wolfgang.duer@uibk.ac.at}
\affiliation{Universit\"at Innsbruck, Institut f\"ur Theoretische Physik, Technikerstra{\ss}e 21a, 6020 Innsbruck, Austria}

\date{\today}

\begin{abstract}
We propose a quantum computation architecture based on geometries with nearest-neighbor interactions, including e.g. planar structures. We show how to efficiently split the role of qubits into data and entanglement-generation qubits. Multipartite entangled states, e.g. 2D cluster states, are generated among the latter, and flexibly transformed via mid-circuit measurements to multiple, long-ranged Bell states, which are used to perform several two-qubit gates in parallel on data qubits. We introduce planar architectures with $n$ data and $n$ auxiliary qubits that allow one to perform $O(\sqrt n)$ long-ranged two-qubit gates simultaneously, with only one round of nearest neighbor gates and one round of mid-circuit measurements. We also show that our approach is applicable in existing superconducting quantum computation architectures, with only a constant overhead.
\end{abstract}

\maketitle
\section{Introduction}
Quantum computers hold the promise to solve problems that are not accessible for classical computing devices, ranging from material design over optimization in logistics to quantum chemistry. It is beneficial to have quantum computers with a large number of qubits, which need to be properly designed in order to allow for an efficient implementation of large-scale quantum computations. Several physical realizations, such as quantum computers based on superconducting qubits \cite{RN26,RN27,RN28}, but also approaches based on trapped atoms \cite{RN29,RN30,RN31,RN32} or ions \cite{RN19,RN20,RN21,RN22,RN23}, can at the moment only realize quantum interactions at a short range, such as nearest-neighbor quantum interactions in a one-dimensional or two-dimensional spatial arrangement of the qubits. While in principle this leaves the computational complexity of the quantum algorithms unaltered, such geometrical restrictions lead to a significant (polynomial) overhead to obtain long-range quantum gates or an all-to-all connectivity between the qubits. Constant factors or poynomial overheads are usually disregarded in complexity assessments, however they are relevant when taking devices of limited size and quality into account. For practical purposes, it is hence beneficial to minimize overheads, and find efficient ways to realize long-distance gates also in such restricted settings.

Here we propose a measurement-assisted quantum computer architecture that utilizes multipartite entangled states, which overcomes limitations posed by architectures based on short-ranged or even nearest neighbor connectivity among the qubits. Our approach does not depend on any physical realization of the qubits, i.e. is platform independent. We show how to split up the qubits into data qubits for quantum information processing, and auxiliary qubits for entanglement generation for different structures to optimize performance. This includes proposals for existing geometries like hexagonal structures currently used in IBMQ quantum processors, but also optimized 2D and 3D architectures. A multipartite entangled state generated among the auxiliary qubits is used as a resource to realize two-qubit and multi-qubit gates on the data qubits by performing suitable quantum protocols that include measurements.

We demonstrate how one can efficiently implement $O(\sqrt n)$ long-range two-qubit gates in parallel among substantially arbitrary pairs of qubits, with only constant overhead, where the total number of qubits is $n$. This can be achieved by performing mid-circuit single-qubit measurements on a 2D cluster state (or more generally on some other multipartite entangled quantum state) that is generated among the auxiliary qubits by utilizing the available short-ranged interactions. The potential power of mid-circuit measurements for state preparation and computation have been investigated in multiple works, see e.g. \cite{RN12,Baeumer2024a,PhysRevApplied.23.014057,Plathanam2025,Baeumer2024,RN9,RN13,RN14,RN15,RN16,RN17,RN18,yang2023,PRXQuantum_6_010306,choe2024}. The mid-circuit measurements allow for an efficient, constant depth implementation of each long-range quantum gate by generating a Bell pair \cite{RN12,Baeumer2024a,PhysRevApplied.23.014057,Plathanam2025}. This is in contrast to gate-based approaches that utilize multiple rounds of SWAP gates to bring qubits next to each other, and which have an overhead that scales with the distance between two qubits, i.e. $O(n)$ in a 1D setting, and $O(\sqrt n)$ in a 2D setting. Previously discussed schemes concentrate on a measurement-assisted implementation of a single long-ranged gate, and the implementation of multiple gates $N$ needs to be done sequentially, i.e. in $N$ steps.
Here we show how a large number $N$ of long-range quantum gates can be realized in such a measurement-assisted manner in parallel in a {\em single} step. To this aim, we utilize the zipper scheme of Ref. \cite{Freund2024} that is based on the X-protocol  of \cite{Hahn2019}, where a 2D cluster state can be flexibly transformed via single-qubit measurements into the desired target configuration of Bell states. In addition, also multi-qubit gates (acting on three or more qubits) or even entire quantum circuits, such as Clifford circuits, can be flexibly implemented using the present approach. Finally, optimal lattice structures are identified to achieve this aim, where the qubits are split into data processing qubits and auxiliary qubits to generate (multipartite) entanglement.

The main contributions of this work can be summarized as follows:
\begin{itemize}
\item We propose a quantum computational architecture that combines gate-based and measurement-based elements. Qubits are split into data qubits and auxiliary qubits, where the latter are used to generate and store (multipartite) entangled states.
\item We show how to overcome the limitation of short-ranged interactions within such a set-up, and realize multiple long-ranged gates in parallel by generating a multipartite entangled state via short-ranged interactions in the auxiliary system, and using it as a resource to perform gates between distant data qubits.
\item We put forward explicit implementations of the scheme for different set-ups, including existing architectures with nearest-neighbor interaction geometries in different 2D lattices.
\end{itemize}

The paper is structured as follows. In Sec. \ref{Sec_Architecture} we describe our architecture, and review how resource states can be manipulated by means of local measurements to obtain different target states, including multiple Bell pairs shared among different parties, and general graph states. In Sec. \ref{Sec_remotegates} we describe how such auxiliary entangled states can be used to perform remote two- and multi-qubit gates. In Sec. \ref{Sec_planar} we consider different examples of planar lattices with nearest neighbor interactions, and show how our approach can be realized in different settings. Sec. \ref{Sec_features} includes a comparison of different lattice structures, and a discussion of the advantages of our approach over other strategies. We discuss the influence of noise and imperfections in Sec. \ref{Sec_Fid}, where we also provide gate fidelities for different scenarios. We summarize and conclude in Sec. \ref{Sec_summary}.

%{\it Setting and scheme.---}
\section{Architecture and resource-state manipulation}\label{Sec_Architecture}
We consider a quantum computer architecture based on a set of qubits that can interact via nearest-neighbor couplings indicated by edges in a given lattice structure. The qubits are divided into data qubits that are used to store and process quantum data, and auxiliary qubits for entanglement generation. See Fig.~\ref{Fig_rectangular_lattice}, showing the example of a quantum system involving a set of data qubits (green) and auxiliary qubits (red) arranged according to a two-dimensional lattice. The lines connecting the qubits in Fig.~\ref{Fig_rectangular_lattice}(a) illustrate the pattern of the available (two-qubit) quantum interactions between the qubits, i.e the auxiliary qubits can interact by means of nearest-neighbor interactions according to the two-dimensional lattice geometry. Further, each data qubit can interact with (at least) one auxiliary qubit along a diagonal edge.
Single-qubit gates are performed directly on the data qubits, while two-qubit and multi-qubit gates acting on multiple data qubits will be implemented by utilizing entanglement shared between the auxiliary qubits in the form of an entangled quantum state \cite{rierasabat2024}.

Here we consider the example of a two-dimensional (2D) cluster state \cite{Briegel2001} on $n$ auxiliary qubits, defined as
$|C_{\rm 2D}\rangle = \prod_{(i,j)\in E}CZ_{ij}|+\rangle^{\otimes n}$. 	
Therein, CZ$_{ij}$ is a controlled-Z gate (“CZ gate” for short) acting on auxiliary qubits $i$ and $j$, and $E$ is the edge set of the two-dimensional lattice of auxiliary qubits, wherein the edge between qubits $i$ and $j$ is denoted as $(i,j)$. The CZ gate is defined by CZ=diag$(1,1,1-1)$, and we denote $|+\rangle=1/\sqrt{2}(|0\rangle + |1\rangle)$. In other words, the 2D cluster state is obtained by applying a product of CZ gates, one for each edge of the two-dimensional lattice of auxiliary qubits, to the initial state $|+\rangle^{\otimes n}$. Notably, the CZ gates respect the pattern of available interactions between the auxiliary qubits, and as they mutually commute, all CZ gates can be performed in parallel, i.e. in a single time step. The 2D cluster state is a special instance of a graph state \cite{RN7,heinEntanglementGraphstates}, where the interaction pattern corresponds to an underlying graph $G$ described by $n$ vertices and a set of edges $(i,j) \in E$.

Fig.~\ref{Fig_rectangular_lattice}(b) illustrates a 2D cluster state that has been prepared on the quantum system of Fig.~\ref{Fig_rectangular_lattice}(a). In Fig.~\ref{Fig_rectangular_lattice}(b), the dashed vertical and horizontal lines indicate the pattern of CZ gates that are applied to the respective auxiliary qubits in order to generate the 2D cluster state, i.e. the entanglement topology. Diagonal solid lines indicate available interactions between data and auxiliary qubits.
With a single round of suitable single-qubit measurements, the 2D cluster state can be flexibly transformed into different target configurations, including multiple, long-distance Bell pairs. It will be useful for the following to describe the action of a (single-qubit) Pauli measurement on a qubit of a cluster state, and more generally any graph state.

Consider an arbitrary graph state, i.e. the graph $G$ is arbitrary. If a qubit of the graph state is measured in the Pauli $X$, $Y$ or $Z$ basis, the resulting quantum state is (up to a local unitary operation) again a graph state, with a different graph \cite{heinEntanglementGraphstates}. Specifically, the action of a Pauli $X$, $Y$ or $Z$ measurement of a qubit located at a vertex $a$ of the graph $G$ can be described, up to local unitary correction operations, by the following graphical rules:
	(i) a $Z$-measurement erases all edges connected to $a$;
	(ii) a $Y$-measurement first inverts the neighborhood graph of $a$ (the neighbors of a are all vertices connected to a by an edge; the neighborhood graph of a is the subgraph existing between the neighbors of $a$; inverting the neighborhood graph of $a$ means that any edge between two neighbors of a is replaced by a non-edge and vice versa) followed by a deletion of all edges connected to $a$;
	(iii) an $X$-measurement first inverts the neighborhood graph of an (arbitrary) neighbor $b$ of $a$, followed by an inversion of the neighborhood graph of $a$, followed by a deletion of all edges connected to $a$, followed by another inversion of the neighborhood graph of $b$.
As will be shown in more detail below, the above graphical rules can be used to show how Bell states can be prepared between pairs of auxiliary qubits by performing single-qubit measurements on a cluster state. The auxiliary qubits may be far away from each other, in other words the Bell states can establish long-range entanglement across the lattice of auxiliary qubits. Bell states prepared in this manner are illustrated in Fig.~\ref{Fig_rectangular_lattice}(c) by the dashed lines connecting pairs of auxiliary qubits.

Each Bell state corresponds to an auxiliary entangled state allows one to implement a two-qubit gate, such as a CNOT or CZ gate, remotely on data qubits that are coupled to the auxiliary qubits of the Bell state, by performing a measurement on the qubits of the Bell state (teleportation-like protocol) \cite{rierasabat2024,Baeumer2024,PhysRevApplied.23.014057}. Accordingly, long-range quantum gates between data qubits can be realized. Notice that the afore-mentioned measurements can be performed simultaneously, so with only one round of $2n$ nearest neighbor gates, and one round of (mid-circuit) measurements one can perform multiple long-range two-qubit gates in parallel. Together with single-qubit operations on data qubits, one can perform multiple rounds of single- and multiple, long-range two-qubit gates to realize arbitrary quantum circuits, overcoming limited connectivity.

We emphasize that the usage of 2D cluster states in the present method is different from measurement-based quantum computation \cite{Raussendorf2001,Raussendorf2003,Briegel2009}, since quantum information is not directly processed by means of multiple rounds of single-qubit measurements, but the 2D cluster state serves as a flexible resource to generate multiple bipartite entangled states, which in turn are used to implement (multiple) long-range gates among data qubits.

\subsection{Manipulation of 2D cluster state - Bell state generation}
A possible scheme \cite{Raussendorf2001,Raussendorf2003,Briegel2009} for establishing a Bell pair (“auxiliary entangled quantum state”) between any two qubits $i$ and $j$ in a 2D cluster state is as follows. One identifies a path of edges from the vertex $i$ to the vertex $j$ (“measurement path”) and performs a $Y$-measurement on all qubits of the path, except the end vertices $i$ and $j$. Further, a $Z$-measurement is performed on each qubit that is a neighbor of at least one vertex of the path. The latter $Z$-measurements disconnect the path from the remainder of the graph (see graphical rule (i) above). It may be envisaged that the $Z$ measurements are performed first, thus establishing a one-dimensional (1D) cluster state of the qubits of the path. The $Y$-measurements, which may be performed thereafter, measure out the qubits of the 1D cluster state except for the endpoints $i$ and $j$, thereby providing a Bell state between the qubits $i$ and $j$ (this can be verified by repeatedly applying the graphical rule (ii) above). We note that this order of the operations, while useful for obtaining an understanding of how a Bell state may be prepared, is not necessary: the $Y$ and $Z$ measurements may be executed in any order, and even simultaneously.

The above-described procedure for preparing a Bell state cuts out a (quasi-) one-dimensional region of the 2D cluster state, which becomes disentangled from the remaining auxiliary qubits. This may hinder establishing further Bell states, e.g. it may limit the possibility of preparing a further Bell state between qubits that lie on opposite sides of said (quasi-) one-dimensional region, since a further measurement path that would cross the disentangled region cannot be used to prepare a Bell state.

In the following, we describe an alternative scheme is to prepare multiple Bell states in arbitrary spatial configurations, which is applicable even if the various measurement paths would cross. The scheme follows the zipper scheme of Ref. \cite{Freund2024} utilizing the X-protocol of \cite{Hahn2019} and is illustrated in Fig.~\ref{Fig_zipper}(a)-(b).
Fig.~\ref{Fig_zipper}(a) shows auxiliary qubits that were initially prepared in a 2D cluster state. It is considered to prepare a Bell state (“auxiliary entangled quantum state”) between two auxiliary qubits (blue) by performing single-qubit operations on the 2D cluster state. A diagonal path between the two auxiliary qubits is considered, wherein the path has a zig-zag diagonal shape consisting of alternating horizontal and vertical edges. Each qubit on the path (except the endpoints) is measured in the $X$ basis \cite{Hahn2019}. Further, some of the qubits adjacent to the endpoints are measured in the $Z$ basis, following the scheme of \cite{Hahn2019}. By performing the $X$-measurements as indicated in Fig.~\ref{Fig_zipper}(a), a Bell state between the two auxiliary qubits (blue) is prepared (this may be verified by repeatedly applying the graphical rule (iii) above). This is the X-protocol as described in \cite{Hahn2019}. Further, the entanglement structure of the remaining quantum state (i.e. the state of the unmeasured auxiliary qubits) is substantially preserved, as was observed and utilized in \cite{Freund2024} in the context of a quantum databus. As shown in Fig.~\ref{Fig_zipper}(a) by the dashed lines, the quantum state obtained after the $X$ measurements is again a graph state having edges that bridge the measurement path (zipper scheme \cite{Freund2024}). What is more, the graph state in question is again a 2D cluster state, up to minor modifications arising from the $Z$ measurements in the vicinity of the endpoints of the path. The measurement path shown in Fig.~\ref{Fig_zipper}(a) is an entanglement-structure-preserving measurement path as described herein. One can hence continue in the same way, and establish further Bell pairs from the remaining entangled state. In particular, one can establish Bell pairs whose paths (in the original lattice) cross, as indicated in Fig.~\ref{Fig_zipper}(b).

Figs.~\ref{Fig_zipper} show examples of (entanglement-structure-preserving) measurement paths that extend along a diagonal of the 2D lattice. The method is not limited to this particular situation. Any two vertices (also called nodes) in a 2D lattice can be connected by a path involving two diagonal portions extending in different directions, with just one turning portion (although paths with several turning portion are also permitted). Each diagonal portion has a zig-zag structure, similar to the path shown in Fig.~\ref{Fig_zipper}(a). Further, the measurements used to generate a Bell state for such a path are similar to the measurements shown in Fig.~\ref{Fig_zipper}(a). Namely, an $X$ measurement is performed on all qubits of the path except for the two endpoints of the path. Further, $Z$ measurements are performed in the areas surrounding the endpoints, as also shown in Fig.~\ref{Fig_zipper}(a). In addition, the qubits in the neighborhood of the turning portion of the path are also measured in the $Z$ basis to disconnect them from the path. The turning portion is the portion of the path where the path changes direction.

By performing the aforementioned $X$ and $Z$ measurements, a Bell pair can be generated between the endpoints of the path, while essentially preserving the entanglement structure of the cluster state. Accordingly, multiple Bell states can be generated between essentially arbitrary pairs of auxiliary qubits, where the measurement paths may or may not cross. Only at the end points and the single turning portion (if present), a small hole is introduced in the lattice when generating Bell pairs in this way, which however does not hinder the generation of additional Bell pairs. Each path has length $O(\sqrt{n})$ , and one can achieve $O(\sqrt{n})$ Bell pairs between distinct pairs of nodes \cite{Freund2024}.

\subsection{Graph state generation from 2D-cluster states}
Also other multipartite states, in particular GHZ states and graph states corresponding to different graphs, can be generated from a 2D cluster state by single-qubit operations, as we now discuss.
This problem is considered in Ref. \cite{Freund2025}, where different methods to are presented.

One particular method involves utilizing the fact that a 2D cluster state is a universal resource for measurement-based quantum computation. This implies that any quantum state, in particular any graph state $|G\rangle$, can be generated deterministically up to local Pauli corrections by means of single-qubit measurements (and the local Pauli corrections, being single-qubit operations, can subsequently be performed by acting directly on the qubits in question). An arbitrary graph state of $m$ qubits can be generated from a 2D cluster state of size $O(m) \times O(m^2)$ where the quadratic scaling reflects the fact that $O(m^2)$ CZ gates (corresponding to edges), together with $O(m^2)$ SWAP operations are sufficient to prepare any graph state. The graph state $|G\rangle$ is thereby prepared on some set of auxiliary qubits.
The zipper scheme can be used to transport (teleport) the auxiliary qubits of the graph state to arbitrary locations in the (partially processed) 2D cluster state, similar to what was described above in relation to the preparation of multiple Bell pairs using the zipper scheme. Accordingly, a graph state on any desired set of auxiliary qubits can be realized by processing a cluster state by single-qubit operations only.

The generation of graph states from cluster states is not restricted to this particular method. For several graph states, more efficient methods exist. For example, a three-qubit GHZ state can be generated as follows. Suppose we wish to prepare a GHZ state on three auxiliary qubits $A_1$, $A_2$ and $A_3$ of the 2D cluster state. Within the cluster state, a path between the first auxiliary qubit $A_1$ and the second auxiliary qubit $A_2$ is identified, and a Bell pair is established between these two qubits using the methods described above (e.g. zipper scheme). The path is continued to the third auxiliary qubit $A_3$, where again entanglement between the second auxiliary qubit $A_2$ and the third auxiliary qubit $A_3$ is established by measuring qubits among the path. The generated state is a 3-qubit graph state with an edge between qubits $A_1$ and $A_2$ and another edge between qubits $A_2$ and $A_3$, which is equivalent to a GHZ state up to local unitary operations (where the qubit $A_2$ is the root qubit of the GHZ state).

Larger GHZ states and larger graph states can be generated via a merging procedure. Consider qubit 1, which is part of a first graph state, with neighbors $N_1$, and qubit 2, which is part of a second graph state, with neighbors $N_2$. Consider an auxiliary qubit $C$, and a first path between qubit 1 and $C$, and a second path between qubit 2 and $C$. Qubit 1 is connected to qubit C by measuring among the first path (zipper scheme), and similarly qubit $C$ is connected to qubit 2 by measuring among the second path (zipper scheme). By performing $Y$-measurements on auxiliary qubit $C$ and qubit 2, qubit 2 is merged into qubit 1. That is, qubit 1 has as neighbors now all its initial neighbors $N_1$ of the first graph, and also all neighbors $N_2$ of qubit 2 of the second graph. For example, one may apply this procedure to the two root qubits of two three-qubit GHZ states, and thereby generate a five-qubit GHZ state.

The generation of arbitrary graph states from a 2D cluster state is discussed in \cite{Freund2025}.

\section{Remote gates from entanglement}
\label{Sec_remotegates}
We now discuss how to utilize Bell states and other graph states to perform remote two-qubit and multi-qubit gates.

\subsection{Remote two-qubit gates via auxiliary entanglement}
Once a Bell state $|B\rangle = CZ|+\rangle|+\rangle = 1/\sqrt{2}(|0\rangle|+\rangle + |1\rangle|-\rangle)$ has been prepared on a pair of auxiliary qubits, it may be used to realize a two-body ate on the associated data qubits by performing a teleportation-like protocol. For example, a CZ gate can be applied to the data qubits. To this end, the two data qubits are coupled to the state $|B\rangle$ via respective CZ gates. Thereafter, the two auxiliary qubits of the Bell pair $|B\rangle$ are measured in the $X$ basis. Depending on the measurement outcomes $(-1)^{m_k}$ with $m_1,m_2 \in \{0,1\}$, local $Z$-correction operations $Z^{m_2}\otimes Z^{m_1}$ on the data qubits are performed. This results in the application of a CZ gate to the two data qubits. This is similar -up to a local basis change- to the remote implementation of gates described in \cite{rierasabat2024,Baeumer2024a}.

It is easy to adapt the above procedure to realize, for example, a CNOT gate acting on the data qubits, since CNOT is equal to CZ up to the application of a single-qubit unitary gates (namely Hadamard gates performed on the target qubit before and after the CZ gate).
Notice that the coupling between the auxiliary qubits and the data qubits can already be done before the 2D cluster state is processed to prepare the Bell state, and hence does not require an additional time step.

In light of the above, a large number $N$ quantum operations, which may be long-range gates, can be realized by processing each of the respective Bell states (and the associated data qubits) in the above-described manner. The corresponding measurement paths may cross in essentially arbitrary ways, and the quantum operations can be performed in parallel.

\subsection{Remote multi-qubit gates and circuits via auxiliary entanglement}
We now show that many-body gates acting on three or more qubits can also be realized using similar procedures.

\subsubsection{Multi-qubit rotations $\exp(-i\alpha Z^{\otimes m})$}
One example involves a family of multi-qubit gates of the form $\exp(-i\alpha Z^{\otimes m})$ acting on $m$ qubits, where m can be any integer larger than one and where the angle $\alpha$ is arbitrary. Such gates can be implemented deterministically using an $m$-qubit GHZ state \cite{rierasabat2024}, defined by
	$|GHZ\rangle =\prod\limits_{j=2}^{m}{C}{{Z}_{1,j}}|+{{\rangle }^{\otimes m}}=1/\sqrt{2}(|0\rangle |+{{\rangle }^{\otimes m-1}}+|1\rangle |-{{\rangle }^{\otimes m-1}})$,
where $|-\rangle = (|0\rangle-|1\rangle)/\sqrt{2}$. That is to say, the state $|GHZ\rangle$, defined on $m$ auxiliary qubits can serve as an auxiliary entangled quantum state that is used for realizing the gate  $\exp(-i\alpha Z^{\otimes m})$ on $m$ associated data qubits. The auxiliary qubits and the data qubits are labelled $A_1$ to $A_m$ and $D_1$ to $D_m$, respectively.

We describe below how the GHZ state can be obtained from a 2D cluster state by performing appropriate single-qubit operations thereon. Before doing so, we show how the GHZ state can be used to realize gates of the form$\exp(-i\alpha Z^{\otimes m})$.
We call the first qubit (i.e. qubit $A_1$) of the GHZ state the root qubit. The gate $\exp(-i\alpha Z^{\otimes m})$ can be realized by coupling the data qubits $D_1$ to $D_m$ to the auxiliary qubits $A_1$ to $A_m$ of the GHZ state, specifically by applying a CZ gate between qubits $D_1$ and $A_1$ and by applying a CNOT gate between qubits $D_j$ and $A_j$, for every $j$ ranging from 2 to $m$. In other words, the overall unitary operator
	$CZ_{D_1A_1}\prod_{j=2}^m CNOT_{D_j,A_j}$
is applied (where it is recalled that the CNOT gates maps $|00\rangle$ to $|00\rangle$, maps $|01\rangle$ to $|01\rangle$, maps $|10\rangle$ to $|11\rangle$ and maps $|11\rangle$ to $|10\rangle)$. Then, the auxiliary qubits $A_1$ to $A_m$ are measured. Specifically, auxiliary qubits $A_2$ to $A_m$ are measured in the $Z$-basis, and a unitary $Z$ correction operation is applied to auxiliary qubit $A_1$ if the parity of the measurement outcomes is odd. This is followed by a single-qubit rotation $e^{-i\alpha X}$ on auxiliary qubit $A_1$, which is subsequently measured in $Z$. A final correction operation $Z^{\otimes m}$ on all data qubits $D_1,\ldots , D_m$ is applied if the result of the last measurement of auxiliary qubit $A_1$ is -1. This realizes the multi-qubit rotation $\exp(-i\alpha Z^{\otimes m})$ on the data qubits $D_1$ to $D_m$.

The above procedure can easily be adapted to realize rotations that are specified by an arbitrary tensor product of Pauli operators rather than $Z^{\otimes m}$, e.g. by applying single-qubit unitary operators to the data qubits in order to rotate Z into X or Y, as desired. Furthermore, sequences of multi-qubit rotations $\exp(-i\alpha Z^{\otimes m})$ applied on different subsets of data qubits, with potentially different angles $\alpha$, allow one to implement arbitrary multi-qubit diagonal gates. A particular interesting instance of gates that can be realized in this way are multi-qubit Toffoli gates \cite{rierasabat2024}(where the number of qubits on which the Toffoli gate acts is arbitrary), which are - up to local unitary operations, which can be easily accounted for - equivalent to a diagonal gate with all diagonal entries being 1, except at the last position which is -1.
Further, it is noted that single-qubit operations can be applied directly to the data qubits in an arbitrary manner. Accordingly, based on the aforementioned gates that can be realized using the GHZ state, many additional gates can be obtained by allowing the gates to be preceded and/or followed by single-qubit gates. Thus, many different multi-qubit gates can be realized by the procedure described above.

\subsubsection{Clifford circuits}
Another example of a family of multi-qubit operations that can be realized using the present method is the family of Clifford operations $U$ acting on an arbitrary number $m$ of qubits \cite{rierasabat2024}. Therein, $m$ may even scale with the system size, i.e. $m$ need not be a constant.  A Clifford operation is any unitary operation from the Clifford group, i.e. any unitary operation that can be obtained as a quantum circuit consisting of CNOT, Hadamard and Phase gates (where the Phase gate is a single-qubit gate mapping $|0\rangle$ to $|0\rangle$ and $|1\rangle$ to $i|1\rangle$, where $i$ is the imaginary unit. Via the Choi-Jamiolkowski isomorphism, there exists for each Clifford operation $U$ on $m$ qubits a corresponding $2m$-qubit state $|{{\psi }_{U}}\rangle $ that is - up to local unitary operations - equivalent to a graph state. The state $|{{\psi }_{U}}{{\rangle }_{AA'}}$ is defined as $U$ acting on system $A\acute{\ }$ (one half of the total system) of a maximally entangled state $\otimes _{j=1}^{m}|{{\phi }^{+}}{{\rangle }_{{{A}_{j}}A{{\text{ }\!\!\acute{\ }\!\!\text{ }}_{j}}}}$ with $|{{\phi }^{+}}\rangle =(|0\rangle |0\rangle +|1\rangle |1\rangle )/\sqrt{2}$.
The state $|{{\psi }_{U}}\rangle $ allows one to implement the Clifford operation $U$ deterministically on an arbitrary input state of the data qubits. This is achieved by teleporting the data qubits through the state $|{{\psi }_{U}}\rangle $, i.e. by performing Bell measurements on each pair consisting of a data qubit $D_j$ and an associated auxiliary qubit $A_j$. Since the operation $U$ is Clifford, Pauli correction operations depending on the measurement outcomes can be performed, yielding a deterministic implementation of $U$. Finally, the output qubits (i.e. the qubits in subsystem $A\acute{\ }$) are swapped with the data qubits to restore the quantum state to its original position \cite{rierasabat2024}.

Further, it is again noted that single-qubit operations can be applied directly to the data qubits in an arbitrary manner, so that additional gates can be realized by allowing the Clifford operations to be preceded and/or followed by single-qubit gates.
We remark that in principle one could also implement single-qubit (Clifford) gates on data qubits via measurements performed on the auxiliary qubit that is entangled with the data qubit. However, we do not consider this possibility here, since (Pauli) correction operations need to be performed on the data qubits anyway, and it seems overly complicated to first entangle the data qubit with the auxiliary qubit, rather to perfrom a desired single-qubit unitary operation on the data qubit directly.

The multipartite states used to realize the aforementioned multi-qubit gates, namely the GHZ state and the state $|{{\psi }_{U}}\rangle $, can be generated from a 2D cluster state by processing the latter by single-qubit operations. We note that both the GHZ state and the state $|{{\psi }_{U}}\rangle$ are stabilizer states, and hence equal to a graph state up to the application of a local unitary operation. Since local unitary operations can be realized by directly acting on the qubits, the methods to generate arbitrary graph states discussed above by processing a cluster state by single-qubit operations suffice.

\section{Examples for nearest-neighbor interaction patterns in planar lattices}
\label{Sec_planar}
In the following, several examples of quantum computation architectures are discussed which can be used to realize the above-described method. We note, however, that also in other underlying planar lattices, our strategy consists in obtaining an entanglement structure that corresponds to the 2D cluster state on a regular rectangular lattice. The reason is that the zipper scheme is only known to work in such a lattice, while the only known strategy to generate Bell pairs in other lattice structures is the $Y$-protocol that also involves measurements of all neigboring qubits among a path. In 2D lattices, the resulting crossings hinder a parallel generation of multiple Bell states.

\subsection{Example 1: Rectangular lattice with open leaves}
An architecture that is particularly well suited to realize the above scheme is given by a 2D rectangular lattice with nearest neighbor interactions for auxiliary qubits, while each data qubit, provided in the center of each plaquette of the lattice of auxiliary qubits, can be coupled to one of the auxiliary qubits via a “diagonal” interaction, as shown in Fig.~\ref{Fig_rectangular_lattice}(a). This allows for the generation of a 2D cluster states in a single time step (see Fig.~\ref{Fig_rectangular_lattice}(b)), and all data qubits have uniform access to long-distance interactions.

One can modify this architecture by attaching multiple data qubits to each auxiliary qubit (so that, in Fig.~\ref{Fig_rectangular_lattice}(a), each plaquette of auxiliary qubits would contain two or more data qubits, each coupled to the same auxiliary qubit). Only one of these data qubits at a time can access a long-range entangling gate in each step, or alternatively two of these data qubits can interact with each other via the same mediating auxiliary qubit. In the latter case, there will be a hole in the 2D cluster state, which does however not hinder the remaining nodes from generating long-range Bell states and implement gates. One may also allow for more auxiliary qubits or qubits for storage, and produce multiple copies of 2D cluster states that can be consumed on demand to generate multiple gates.

\subsection{Example 2: Triangle-square lattices}
Other planar structures are also suitable to implement the above scheme, as illustrated in Figs.~\ref{Figure_triangle_square_lattice.png}(a)-(c). The underlying lattice is a triangular-rectangular lattice. Fig.~\ref{Figure_triangle_square_lattice.png}(a) shows the pattern of available interactions in such a setting, indicated by the lines between the qubits. As before, data qubits are shown as red and auxiliary qubits are blue. In such a system, a 2D cluster state of the auxiliary qubits can be prepared in a “tilted” orientation, where the edges of the 2D cluster state (corresponding to the application of CZ gates) extend diagonally, as illustrated in Fig.~\ref{Figure_triangle_square_lattice.png}(b). As before, each plaquette of the 2D cluster state includes a data qubit. Further, the 2D cluster state can again be prepared in a single time step. Within this setting, multiple quantum operations can be realized by processing the 2D cluster state to prepare Bell states in essentially the same manner as described above, as illustrated in Fig.~\ref{Figure_triangle_square_lattice.png}(c). We note that other variants of rectangular-triangular lattices than the one shown in Fig.~\ref{Figure_triangle_square_lattice.png} exist, which are also suitable.

\subsection{Example 3: Decorated honeycomb lattice}
We will now turn to an underlying interaction pattern corresponding to a decorated honeycomb lattice \cite{RN26,Baeumer2024}, and discuss two possible asignments of data and auxiliary qubits.

\subsubsection{Dense data qubits}
Existing quantum computers based on superconducting qubits (e.g. the IBMQ quantum computer) may have a limited connectivity, for example in the manner depicted in Fig.~\ref{Figure_decorated_honexcomb_lattice}(a) \cite{RN26,Baeumer2024}.
Fig.~\ref{Figure_decorated_honexcomb_lattice}(a) shows qubits arranged according to a decorated hexagonal lattice (also called decorated honeycomb lattice). The pattern of available interactions is again indicated by the lines between the qubits. Such a system of qubits can also be used to perform the methods described herein, as follows. The qubits are divided into data qubits (green), auxiliary qubits (red), and ancilla qubits (blue).

A 2D cluster state can be prepared on the auxiliary qubits by performing a sequence of (nearest-neighbor) operations as illustrated in Fig.~\ref{Figure_honexcomb_gatesequence}. Fig.~\ref{Figure_honexcomb_gatesequence} shows a plaquette of the decorated hexagonal lattice, where the qubits are numbered from 1 to 9. Further, qubits 4’, 5’ and 6’ are also shown, which stem from a similar numbering of the adjacent plaquette. Qubits 1, 3, 7 and 9 are data qubits (green). Qubits 2, 4, 6, 8, 4’ and 6’ are auxiliary qubits (red). Qubits 5 and 5’ are ancilla qubits (blue).
It can also be seen that nine of the original qubits are used to obtain four data and four auxiliary qubits for entanglement generation. The nearest neighbor decoration qubits to the right are used as data qubits (green), while the remaining decoration qubits (blue) among vertical edges are used to enhance entanglement generation, see Fig.~\ref{Figure_decorated_honexcomb_lattice}(b). Entanglement generation in horizontal direction needs to be done bypassing a data qubit without altering it, which requires SWAP operations. The vertical entanglement edge can be generated utilizing the blue decoration qubits, by generating a GHZ state between the two red and the blue decoration qubit, and measuring the blue decoration qubit. This provides a balance between the available number of data qubits, and efficiency to generate the required 2D cluster state and hence obtain long-ranged gates, see Fig.~\ref{Figure_decorated_honexcomb_lattice}c.

%----

\subsubsection{Sparse data qubits}
The decorated honeycomb lattice can also be utilized in a different way to obtain a 2D cluster state, as shown in Figs.~\ref{Figure_alternative_honeycomb.png}(a)-(d).
Fig.~\ref{Figure_alternative_honeycomb.png}(a) again shows a decorated honeycomb lattice, where the qubits are now divided in a different manner into data qubits (green), auxiliary qubits (red), and ancilla qubits (blue). In this case, one can generate entanglement between a pair of auxiliary qubits, either in the vertical or horizontal direction, via a path of ancilla qubits extending between these auxiliary qubits, by (i) applying CZ gates between neighboring qubits on the path, and (ii) measuring all qubits along the path, except for the two auxiliary qubits at the endpoints, in the $Y$ basis. The procedure is performed in two steps, e.g. first the vertical connections (Fig.~\ref{Figure_alternative_honeycomb.png}(b)), and then the horizontal connections (Fig.~\ref{Figure_alternative_honeycomb.png}(c)). Accordingly, a 2D cluster state can be prepared on the auxiliary qubits. Further, each data qubit can be coupled to an auxiliary qubit (Fig.~\ref{Figure_alternative_honeycomb.png}(d)) in a similar manner, i.e. using a path of ancilla qubits extending from the data qubit to the auxiliary qubit. Notice that this scheme involves a (constant) overhead in the required number of qubits (ten qubits are needed in total for each data and auxiliary qubit). On the other hand, this approach offers the advantage that fewer steps are required to generate the 2D cluster state. Further, the data qubits are left untouched in the entanglement generation process, which avoids introducing additional errors on qubits that do not participate in a two-qubit gate. Entanglement can be generated beforehand and stored until it is needed to perform gates on data qubits in a flexible way.

\subsection{Further examples of planar lattices}

2D cluster states can be generated from graph states corresponding to other lattice structures. For example, by performing suitable single-qubit operations, a graph state corresponding to a honeycomb lattice can be transformed into a graph state corresponding to a triangular lattice, which can in turn be transformed into a graph state corresponding to a Kagome lattice and finally to a 2D cluster state on a coarse-grained lattice \cite{VandenNest2007}. Some of the available qubits that do not participate in the 2D cluster state can be used as data qubits. This shows that other architectures corresponding to different spatial arrangements of the qubits, i.e. different from a rectangular 2D lattice, are suitable for the present approach.

\section{Features of different lattice structures}
\label{Sec_features}

\subsection{3D geometry}
Three-dimensional (3D) architectures based on 3D lattices with e.g. nearest-neighbor connectivity can also be used to establish multiple long-range gates in parallel, similarly to the above-described method that uses a 2D cluster state or other planar structures. For cubic lattices, 3D cluster states can be generated, and likewise utilized to obtain multiple Bell pairs in parallel. A Bell pair can be generated between any pair of qubits of the 3D cluster state by identifying a path between them. All qubits (except the end nodes) of the path are measured in the $Y$ basis, and all neighboring qubits of the path are measured in $Z$. The $Z$ measurements isolate the path - and hence a 1D cluster state - from the rest of the qubits, while the $Y$ measurements establish a Bell pair among the end nodes. Notice that there are (at most) four neighboring qubits for each qubit on the path. Any qubit in a 3D cubic lattice has 6 neighbors - one belongs to the incoming, and one to the outgoing path, and some of the neighbors may also be neighbors of other qubits among the path in case of turning points. In contrast to planar lattices, such a strategy can also be employed for multiple paths and hence the parallel generation of multiple Bell states, as the third dimension can be used to avoid crossings.

3D structures offer certain advantages over 2D structures. First, the maximum distance between any two points in a 3D rectangular lattice is smaller - at most $3 n^{1/3}$, while the distance is up to $2 n^{1/2}$ in a planar lattice, where we assume the same total size of the resource state, $n$, which are arranged as $\sqrt{n}\times \sqrt{n}$ 2D lattice, or a $\sqrt[3]{n}\times \sqrt[3]{n} \times \sqrt[3]{n}$ 3D lattice.

Second, in a 3D system one can obtain more long-range Bell pairs in parallel as compared to a 2D structure. More concretely, one obtains up to $O(n^{2/3})$ Bell pairs, as compared to $O(n^{1/2})$ in a 2D structure. The generation of a single Bell pair cuts essentially a one-dimensional hole into the 3D structure. This, however, still leaves enough room to have more parallel paths. Since the length of each path is $O(\sqrt[3]{n})$, one can have up to $O(n^{2/3})$ parallel pathes that do not intersect.

\subsection{Nested structures for gate routing}
One can also use 2D cluster states at different scales to form a nested structure for long-range entanglement (and hence gate) routing. Such an approach is particularly useful when the data qubits can be organized in groups with different required connectivity (in terms of entangling gates) between groups. Such hierarchical structures are beneficial in network design, and can also be utilized in quantum processor architectures. With the present approach, this translates directly to the entanglement structure of the underlying multipartite entangled state that is generated and further manipulated.

\subsection{Advantages of our architecture and the 2D cluster state}
Here we compare our architecture with other ways to generate long-ranged gates, and specifically discuss the advantage of using the 2D cluster state as a reosurce.

\subsubsection{Comparison to alternative ways of obtaining long-ranged gates}
There are different ways to overcome restrictions of short-ranged couplings. In several platforms, there exist ways of realizing long-ranged gates via long-ranged couplings directly, which can be viewed as a hardware solution. This, however, comes with increased demands on hardware control and design, and in several platforms is simply not applicable.

Our approach provides an alternative and conceptually new way of overcoming these difficulties, which is in principle platform independent and does not require a direct physical long-ranged interaction. It utilizes only nearest-neighbor couplings and mid-circuit measurements, and in this way shifts the requirements towards a high-fidelity implementation of these specific tasks. Our approach can be viewed as a hardware-independent solution.

\subsubsection{Advantage of 2D cluster state over bipartite entanglement}
First, it should be stressed that using mid-circuit measurements in the first place offers advantages over purely gate-based approaches. In a gate-based architecture with nearest neighbor gates, gates between remote qubits are usually realized by means of multiple SWAP operations that transport the qubits so that they are next to each other. In a 2D setting, this requires $O(\sqrt n)$ sequential SWAP gates.

Previous schemes that utilize mid-circuit measurements have concentrated on the realization of a single long-ranged two qubit gate, and only gates where the corresponding connecting pathes do not cross can be realized in parallel. Realizing additional long-ranged gates require a sequential application, i.e. the required number of steps is given by the number of long-ranged gates to be implemented.

Here, in contrast, we use a multipartite entangled state as resource, specifically a 2D cluster state. There are multiple advantages in such an approach, as compared to utilizing only bipartite entanglement (Bell states). First of all, cluster states can be established beforehand, even before it is known which gates should be performed. The cluster state can then be flexibly transformed to the desired target configuration by means of a single round of local measurements. This allows one to perform multiple two-qubit gates in a single step (i.e. a very small computational depth), where it can be essentially freely chosen among which qubits. While also Bell states can in principle be established beforehand, this requires a quadratically larger storage capability if full flexibility without prior knowledge is demanded - there are $n(n-1)/2$ different Bell states.

The zipper scheme allows one to generate multiple Bell pairs in parallel from a 2D cluster state, which leads to a smaller number of qubits that need to be measured as compared to schemes based on cutting paths. Alternatively, the cluster state can also be transformed to other graph states among certain target qubits \cite{Freund2025}, which allows one to perform multi-qubit gates or entire Clifford circuits \cite{rierasabat2024} - again in a flexible way and in a single step.

\subsubsection{Comparison to measurement-based quantum computation}
We also stress that the way we utilize the 2D cluster state as a resource in our scheme is conceptually different to measurement-based quantum computation (MBQC) \cite{Raussendorf2001,Raussendorf2003,Briegel2009}. MBQC corresponds to a different computational model, with distinct features and overheads. In particular, in MBQC there are dedicated input and output qubits, while the rest of the system serves as resource to perform the computation. In MBQC the size of the resource state is given by the length/complexity of the associated circuit that should be implemented. The qubits on the left border of the cluster state are input qubits, and qubits within the bulk of the 2D cluster state are measured in particular basis to process the information. Gates are thereby translated to measurement patterns, and the required size of the cluster state is given by the number of (nearest neighbor) two-qubit gates and single qubit gates that should be implemented. The output of the computation is then at the outermost right part of the 2D cluster state. A 2D cluster state of size $n \times m$ can thereby be used to perform $O(m)$ gates on a register of $n$ qubits initially at one border of the cluster state, and ending up at the other side.

Here, in contrast, we consider a different computational model, where the 2D cluster state is used in a different way. Each qubit of the 2D cluster state is associated with a data qubit, and there are no dedicated input- or output systems. Rather each of the qubits of the 2D cluster state takes an equal role, and can be chosen to take part in a two-qubit gate acting on the associated data qubit. The cluster state is thereby used as a resource to perform specific gates – namely long-ranged two-qubit gates. The manipulation of the cluster state is different than in MBQC. There is no circuit that is mapped, there is rather a direct transformation of the multipartite entanglement to different Bell pairs, which can be shared between basically arbitrary pairs of qubits within the cluster state (and not just on border). In this case, the size of the register is $n \times m$.

\subsection{Comparison to sequential implementation - without generating 2D cluster states}
Finally, we compare our approach to a sequential implementation of gates. In \cite{Baeumer2024a}, a single implementation of a long-ranged CNOT gate is discussed, which requires a constant depth circuit that essentially prepares a 1D cluster state, assisted by mid-circuit measurements. Clearly, this approach allows one to perform also multiple gates in a sequential way, where the required depth is given -without further optimization- by the number of gates that should be performed. However, one can parallelize the implementation by generating multiple Bell states in parallel, where one assigns a path to each desired connection. This is however limited to pathes that do not cross. We now show that one can overcome this limitation, see Fig.~\ref{Fig.~sequential}, without first preparing a 2D cluster state as resource, and subsequently processing this resource state by single-qubit operations as described above.

Fig.~\ref{Fig.~sequential}(a) shows a quantum system including data qubits (green) and auxiliary qubits (red) arranged according to a regular two-dimensional lattice. This is similar to e.g. Fig.~\ref{Fig_rectangular_lattice}(a); yet in the present case, no cluster state or other entangled resource state will be prepared on the auxiliary qubits.
It is envisaged to generate Bell pair between some of the auxiliary qubits, as indicated in Fig.~\ref{Fig.~sequential}(c) by the dashed lines between pairs of auxiliary qubits. This will be achieved by performing single-qubit measurements along paths connecting the auxiliary qubits in question.

In a first round, as illustrated in Fig.~\ref{Fig.~sequential}(a), all non-conflicting connections of the paths are established by performing nearest-neighbor CZ gates (indicated by solid vertical and horizontal lines between the auxiliary qubits in question). At crossings between two paths, only the horizontal connections are established. This is followed by $Y$-measurements on all qubits of each path, except for the end points of the path, resulting in multiple Bell pairs being generated, as indicated in Fig.~\ref{Fig.~sequential}(b) by the dashed lines. Notice that each path may be intersected at multiple crossings.
In a second round, as further illustrated in Fig.~\ref{Fig.~sequential}(b), the vertical connections at each crossing are established by performing the appropriate CZ-gates (solid vertical lines in Fig.~\ref{Fig.~sequential}(b)), followed again by $Y$-measurements on all intermediate qubits. As a result, multiple long-range Bell pairs are obtained as illustrated by the dashed lines in Fig.~\ref{Fig.~sequential}(c), which can be used to implement long-range gates among the data qubits in question.

With the above approach, $O(\sqrt n)$ Bell pairs can be established in parallel by using only two rounds of nearest neighbor gates, and two rounds of mid-circuit measurements, which allows to implement multiple long-range gates. Bottlenecks only appear at intersection points of two paths, where each crossing can be resolved in two rounds. Overall, this procedure is less efficient (by a factor of two) than the approach proposed above based on an entangled resource state such as a cluster state.

\section{Gate fidelities}
\label{Sec_Fid}
Here we briefly discuss the relation between nearest-neighbor physical gate fidelities, lattice geometry, and the quality of the realized long-range gates (see also \cite{RN12,Baeumer2024a,PhysRevApplied.23.014057}).
One can use the noisy stabilizer formalism \cite{MorRuiz2023} to efficiently compute the fidelity of long-distance Bell pairs generated from 1D or 2D cluster states, taking noise in the preparation of the 2D cluster state and imperfections in measurements into account. Fidelities of Bell pairs generated in 2D (and other) lattice structures subjected to different kinds of Pauli noise modeling imperfect state preparation, storage and measurements, were computed in \cite{MorRuiz2023}. It was shown that the fidelity is determined by number of measured qubits, which is given by the length of the path and the number of neighbors that need to measured when using a cutting strategy. We note that using the zipper scheme (involving $X$ measurements) to generated Bell pairs by measuring along diagonal paths reduces the number of required measurements, and hence increases fidelity. The fidelity of Bell pairs directly translates into gate fidelity of long-ranged gates.
This also implies that in 3D structures, larger gate fidelities can be achieved since fewer measurements are required to generate a long-ranged Bell pairs for large system sizes \cite{MorRuiz2023}.

We assume a single qubit depolarizing noise acting on each of the qubits of the resource state, where noise on qubit $a$ is described by ${\cal E}_a\rho = p \rho + \tfrac{1-p}{4} \sum_{i=0}^{3} \sigma_i^{(a)} \rho \sigma_i^{(a)}$. This noise reflects imperfect resource state preparation, but can also be used to take imperfections in measurements and operations performed on each of the qubits into account. Neglecting details of the particular kind of resulting noise, and assuming small error probabilities $\epsilon$ (i.e. $p \approx 1-\epsilon$), it follows from \cite{MorRuiz2023} and \cite{MorRuiz2024} that the fidelity of the resulting Bell state generated from a 2D cluster state using the X-protocol is given by $F \approx 1-1/2(n_{\rm 2D}+9)$, where $n_{\rm 2D} = O(\sqrt n)$ is the length of the path in a 2D setting. In 3D lattice, the resulting fidelity of a Bell pair generated by the Y-protocol is given by \cite{MorRuiz2024} $F \approx 1-\tfrac{1}{2}(5n_{\rm 3D} +2)+3)\epsilon$, with $n_{\rm 3D} = O(n^{1/3})$ is the length of the path in a 3D lattice. Notice that in contrast to the 2D setting, also neighbors along the path need to be measured, resulting in the factor 5. This fidelity directly translates into the gate fideliy of the associated CNOT or CZ gate, except for the additional influence of local noise on the data qubit.

\section{Summary and outlook}
\label{Sec_summary}
The above-described method involves a quantum computer architecture that enhances limited, nearest neighbor connectivity to full connectivity by means of multipartite entanglement. The key feature is to split the quantum system into data qubits and auxiliary qubits, where a 2D cluster state, or another multipartite entangled resource state, is generated and shared among the auxiliary qubits. This multipartite entangled state can be flexibly transformed into different target configurations by means of single-qubit measurements, including multiple long-range Bell states shared between distant pairs of qubits. In turn, these Bell states allow one to implement multiple long-range two-qubit gates in parallel. This leads to a quantum processor architecture where multipartite entangled states are utilized to enhance flexibility and long-range connectivity, and measurement-based elements support a circuit-based design.

We point out that the kind of entanglement that is generated and utilized can be adapted to the task and purpose at hand. For example, if only a restricted set of target configuration of Bell states (or long-ranged gates) should be flexibly realized, other resource states with a potentially simpler entanglement structure might be suitable \cite{RN81}. Furthermore, entanglement can be generated beforehand, and in fact other entangled states might be more powerful than a naturally available 2D cluster state, e.g. a 3D cluster states allows one to perform more gates in parallel as we have also illustrated. It might also be interesting to add elements of blind quantum computation to the proposal, i.e. consider techniques and methods to hide the performed protocol or computation from a server that performs the computation.
From a more physical perspective, it would be interesting to optimize the suggested approach to different experimental platforms. While we have concentrated on superconducting quantum processor architectures, also other set-ups are viable to our approach. For instance, dual species neutral atom arrays \cite{RN34} offer a natural separation into data and auxiliary qubits for entanglement generation.

\section*{Acknowledgments}
This research was funded in whole or in part by the Austrian Science Fund (FWF) 10.55776/P36009, 10.55776/P36010 and 10.55776/COE1. For open access purposes, the author has applied a CC BY public copyright license to any author accepted manuscript version arising from this submission. Finanziert von der Europ\"aischen Union - NextGenerationEU. We thank M. Van den Nest for valuable suggestions to improve the presentation of the manuscript. The method described here was filed as a patent.

\bibliographystyle{apsrev4-2}
\bibliography{QC_Architecture_submitt.bib}

\newpage
\begin{widetext}
%\newpage
%\section*{ }
%\newpage

\begin{figure}[ht]
    \centering
    \includegraphics[width=\columnwidth]{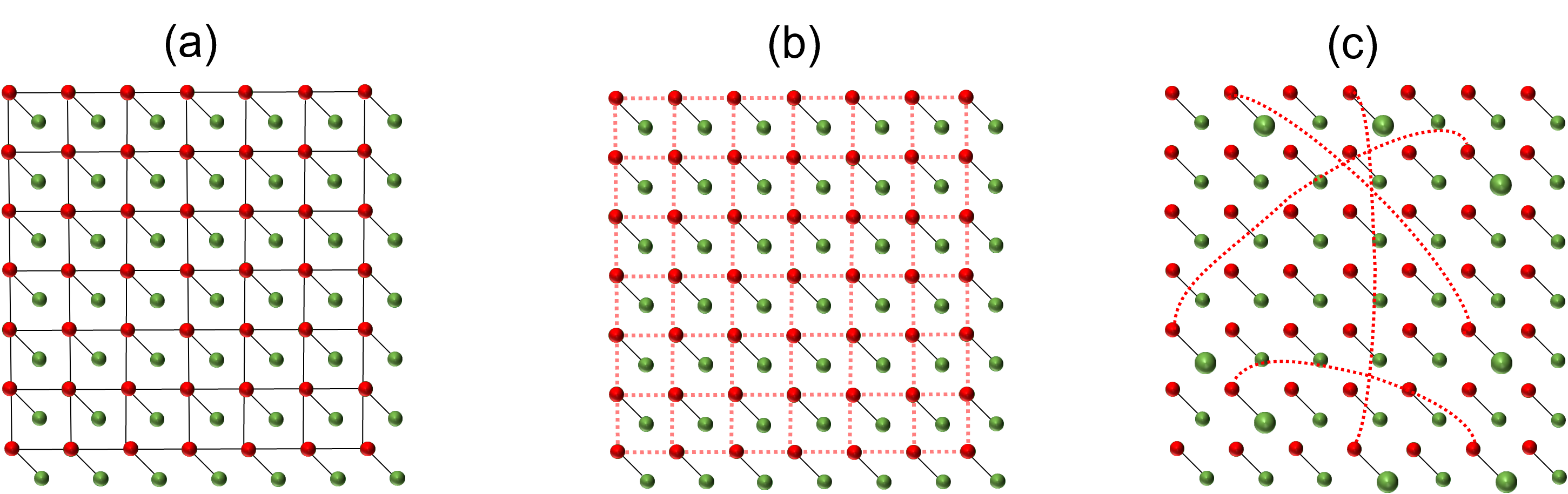}
%\vspace{-9.5cm}
    \caption{\label{Fig_rectangular_lattice} Rectangular lattice with open leaves: (a) shows the connectivity pattern, and the separation into data qubits (green) and auxiliary qubits (red). Multipartite entanglement (indicated by dashed lines) in the form of a 2D cluster state is generated between auxiliary qubits (red) using only nearest-neighbor gates (see (b)), and manipulated by means of suitable measurements using the zipper scheme to generate multiple Bell pairs (c). The Bell pairs are used to implement multiple long-ranged two-qubit CZ gates on data qubits (green).}
\end{figure}

%\end{widetext}

\begin{figure}[ht]
    \centering
    \includegraphics[width=0.5\columnwidth]{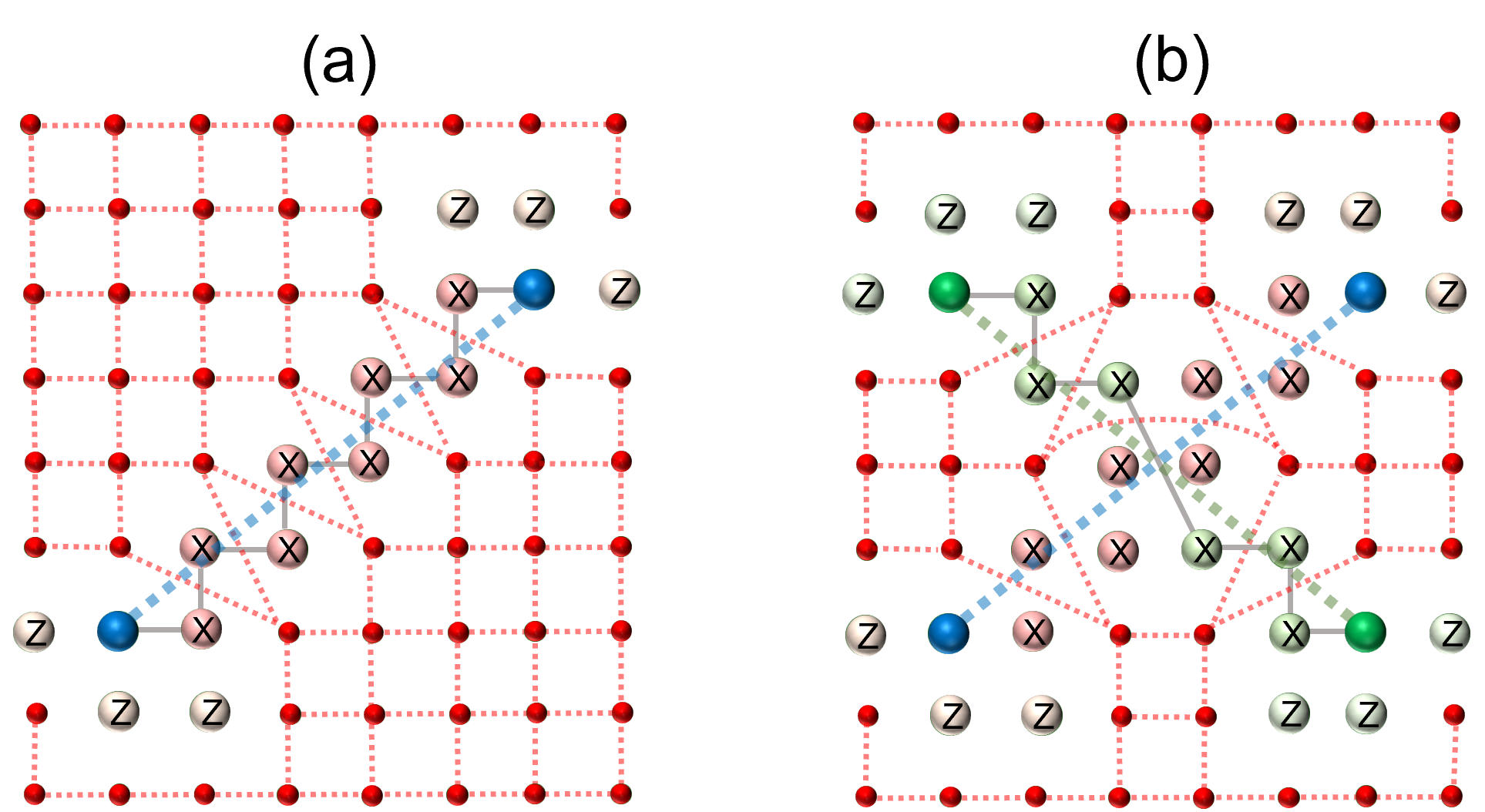}
%\vspace{-1cm}
  \caption{\label{Fig_zipper} (a) Zipper scheme to generate entanglement among a diagonal path by performing $X$-measurements on a 2D cluster state. The remaining entanglement structure, indicated by red dotted lines, closes over the hole. This can be checked by applying the rule for $X$-measurements sequentially on qubits among the path, with the remaining neighboring vertex chosen as special neighbor. (b) Second Bell pair is generated from the remaining structure, where the connecting pathes cross in the original 2D lattice. Around end-nodes, some qubits need to be measured in $Z$ to disconnect the Bell pair from the remaining state.
    }
\end{figure}

%\begin{widetext}

\begin{figure}[ht]
    \centering
    \includegraphics[width=\columnwidth]{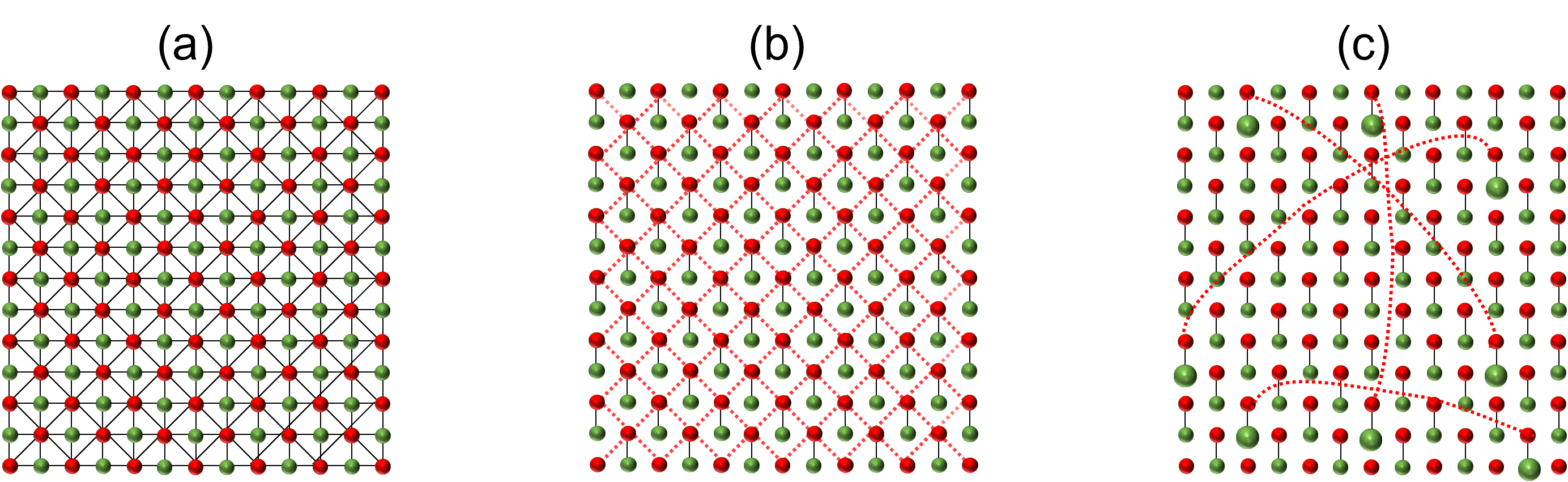}
%\vspace{-10cm}
 \caption{\label{Figure_triangle_square_lattice.png} Triangle-square lattice: (a) shows the connectivity pattern and the separation into data qubits (green) and auxiliary qubits (red). Multipartite entanglement (indicated by dashed lines) in the form of a 2D cluster state is generated between auxiliary qubits (red) (see (b)), and manipulated by means of suitable measurements using the zipper scheme to generate multiple Bell pairs (c). The Bell pairs are used to implement multiple two-qubit CZ gates on data qubits (green).}
\end{figure}

\begin{figure}[ht]
    \centering
    \includegraphics[width=\columnwidth]{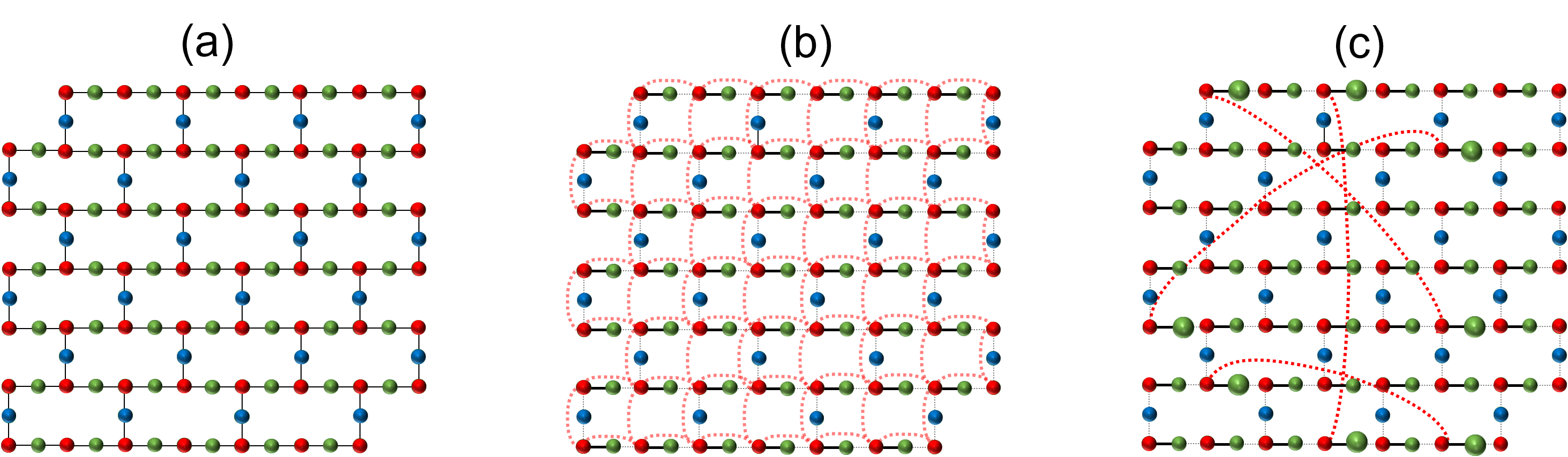}
%\vspace{-10cm}
  \caption{\label{Figure_decorated_honexcomb_lattice} Decorated honeycomb lattice: (a) shows the connectivity pattern corresponding to superconducting quantum processors (IBMQ). The qubits are separated into auxiliary qubits for entanglement generation (red, original honeycomb lattice), data qubits (green) and ancilla qubits (blue) to assist in the entanglement generation, both corresponding to decoration qubits of the lattice. Entanglement (indicated by dashed lines) in the form of a 2D cluster state is generated between auxiliary qubits (red) (see (b)), and manipulated by means of suitable measurements using the zipper scheme to generate multiple Bell pairs (c). The Bell pairs are used to implement multiple two-qubit CZ gates on data qubits (green).}
\end{figure}

%\end{widetext}

\begin{figure}[ht]
    \centering
    \includegraphics[width=0.5\columnwidth]{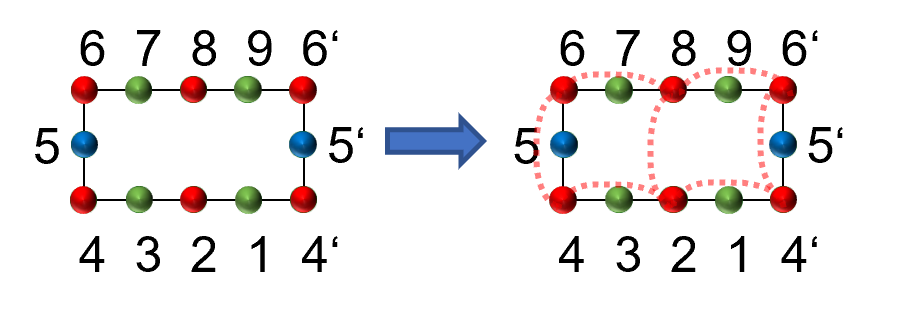}
%\vspace{-1cm}
  \caption{\label{Figure_honexcomb_gatesequence}
  Required steps to generate a 2D cluster state among the auxiliary qubits (red). The following operations are performed (where the auxiliary qubits 2, 4, 6, 8, 4’ and 6’ and the ancilla qubits 5 and 5’ are initially prepared in the state $|+\rangle$, and qubits 4',5',6' are actually also part of the next block, where the same sequence of operations is performed (and hence 4'-6' are entangled)).
 Step 1: Entangle qubits 4 and 5, entangle qubits 5 and 6, SWAP qubits 2 and 3, and SWAP qubits 7 and 8.
 Step 2: measure qubit 5 in the $Y$ basis, which results in qubits 4 and 6 being entangled; also, SWAP qubits 2 and 4 and SWAP qubits 6 and 8.
 Step 3: entangle qubits 2 and 4, entangle qubits 6 and 8, entangle qubits 2 and 5 and entangle qubits 5 and 8.
 Step 4: measure qubit 5 in the $Y$ basis, which results in qubits 2 and 8 being entangled; also, SWAP qubits 2 and 4 and SWAP qubits 6 and 8.
 Step 5: SWAP qubits 2 and 3, SWAP qubits 7 and 8, SWAP qubits 1 and 4’ and SWAP qubits 9 and 6’.
 Step 6: entangle qubits 2 and 4’ and entangle qubits 8 and 6’.
 Step 7: SWAP qubits 1 and 4’ and SWAP qubits 9 and 6’.
 In the preceding steps, the operation “Entangle qubits a and b” means that a CZ gate is applied to the qubits in question. Further, “SWAP qubits a and b” means that a SWAP gate is applied. After performing steps 1 through 6 on all plaquettes, a 2D cluster state is prepared on the auxiliary qubits, as illustrated in Fig.~\ref{Figure_decorated_honexcomb_lattice}(b).
  }
\end{figure}

%\begin{widetext}

\begin{figure}[ht]
    \centering
    \includegraphics[width=\columnwidth]{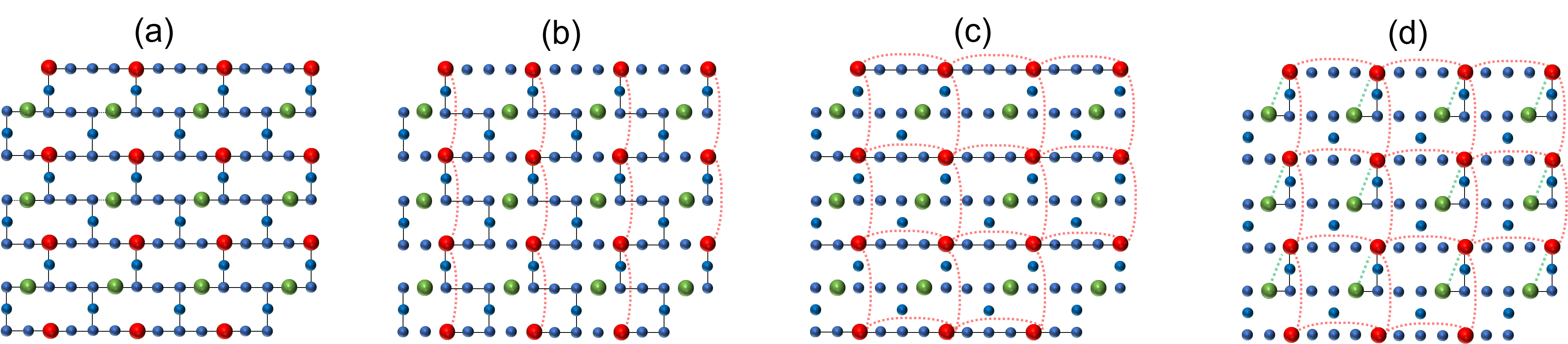}
%\vspace{-1cm}
  \caption{\label{Figure_alternative_honeycomb.png} (a) Decorated honeycomb lattice with alternative association of qubits into data qubits (green), auxiliary qubits for entanglement generation (red) that constitute a 2D cluster states, and additional qubits to mediate entanglement (blue). (b) First vertical connections in 2D cluster state are generated by performing CZ operations as indicated, followed by $Y$-measurements of involved blue qubits. (c) In a second step, horizontal links are established in the same way. (d) Data qubits are coupled to 2D cluster state.
  }
\end{figure}

\begin{figure}[ht]
    \centering
    \includegraphics[width=\columnwidth]{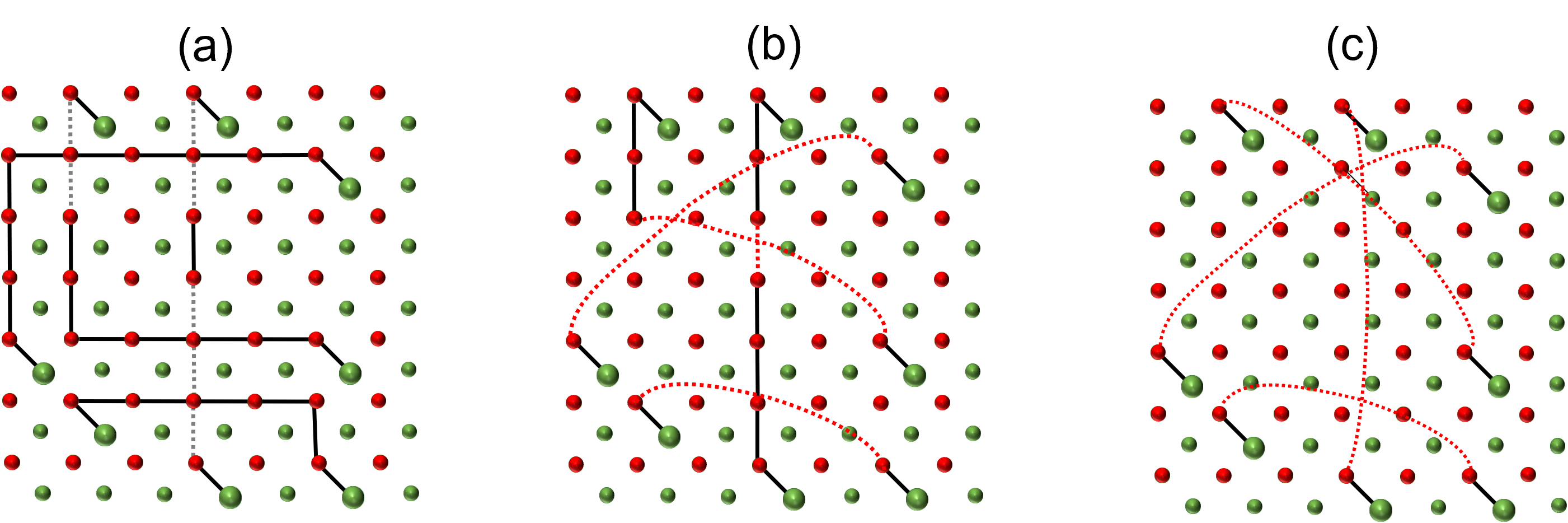}
%\vspace{-1cm}
  \caption{\label{Fig.~sequential} Sequential generation of multiple Bell pairs with connectivity given by a 2D rectangular lattice, see Fig.~\ref{Fig_rectangular_lattice}. (a) In round 1, all non-conflicting connections among each path are established by performing nearest-neighbor CZ gates (indicated by solid lines). At crossings between two pathes, only the horizontal connections are established. This is followed by $Y$-measurements on all qubits, except end points of each line, leading to multiple Bell pairs indicated by red dotted lines. Notice that each path may be intersected at multiple crossings. (b) In round 2, the vertical connections at each crossing are established by performing the appropriate CZ-gates, followed again by $Y$-measurements on all intermediate qubits. (c) This leaves one with multiple long-ranged Bell pairs (red dotted lines), which can be used to implement long-ranged CZ or CNOT gates among the data qubits (marked by enlarged qubits).
  }
\end{figure}

\end{widetext}

\end{document}